\documentclass[a4paper,11pt]{article}
\usepackage{pos}

\title{Search for TeV emission from the Fermi Bubbles at low Galactic latitudes with H.E.S.S. inner Galaxy survey observations}
 \ShortTitle{Search for TeV emission from the Fermi Bubbles with H.E.S.S.}

\author*[a]{Emmanuel Moulin}
\author[a]{Alessandro Montanari}
\author[b]{Denys Malyshev}
\author[c]{Dmitry Malyshev}

\forColl{H.E.S.S.}

\affiliation[a]{IRFU, CEA, Universit\'e Paris-Saclay,
  F-91191 Gif-sur-Yvette, France}
  
\affiliation[b]{Institut f\"ur Astronomie und Astrophysik, Universit\"at T\"ubingen, Sand 1, D 72076 T\"ubingen, Germany}

\affiliation[c]{Friedrich-Alexander-Universit\"at Erlangen-N\"urnberg, Erlangen Centre for Astroparticle Physics, Erwin-Rommel-Str. 1, D 91058 Erlangen, Germany}

\emailAdd{emmanuel.moulin@cea.fr}
\emailAdd{alessandro.montanari@cea.fr}
\emailAdd{denys.malyshev@astro.uni-tuebingen.de}
\emailAdd{dmitry.malyshev@fau.de}

\abstract{The Fermi Bubbles were discovered about a decade ago in the {\it Fermi}-LAT data as a double-lobe structure extending up to 55° in Galactic latitudes above and below the Galactic Center. At the moment their origin is still unknown. The H.E.S.S. collaboration is currently performing the first ever survey in TeV gamma rays of the Milky Way inner region: the Inner Galaxy Survey. This survey is intended to achieve the best sensitivity to faint and diffuse emissions in a region of several degrees around the Galactic Centre. It provides an unprecedented sensitivity to dark matter signals, new diffuse emissions, and TeV outflows from the Galactic Centre. Understanding the properties of the Fermi Bubbles at low Galactic latitudes will provide key insights into their origin. We search for TeV emission at the base of the Fermi Bubbles using low-latitude
spatial templates. The first results obtained with the 2014-2020 H.E.S.S. observations will be reported.}

\FullConference{37$^{\rm{th}}$ International Cosmic Ray Conference (ICRC 2021)\\
		July 12th -- 23rd, 2021\\
		Online -- Berlin, Germany}

\begin{document}
\maketitle

\section{Introduction}
Fermi Bubbles (FBs) are two large structures extending up to about 55$^\circ$ above and below the Galactic center (GC). They were discovered in {\it Fermi}-LAT data~\cite{Dobler:2009xz,Su:2010qj}. At Galactic latitudes $|b| > 10^\circ$, their morphology is consitent with a uniform distribution and their energy spectrum is $\propto E^{-2}$ with an energy cutoff or a significant softening above $\sim100$ GeV. FBs have possible counterparts at other wavelenghts such as the microwave haze~\cite{Dobler:2009xz,2013A&A...554A.139P} 
and X-ray features near the GC~\cite{Ponti:2019swx} and at higher latitudes~\cite{Predehl:2020kyq}. 

The spatial and spectral characteristics of the FBs can be explained by leptonic and hadronic models, where relativistic protons/electrons are injected into the medium with outflows, continuously or sporadically operating in the past, from the region spatially close to the GC. Such outflows could be produced by the past activity of the supermassive black hole Sagittarius A* at the GC~\cite{Guo:2011eg,Guo:2011ip,Bland-Hawthorn:2013gna,Barkov:2013gda}, star formation close to the GC~\cite{Crocker:2010dg}, or outflows driven by multiple core-collapse supernovae~\cite{2021ApJ...913...68Z}.
A recent analysis of the {\it Fermi}-LAT data at low Galactic latitudes ($|b| < 10^\circ$) suggests that the intensity of the gamma-ray emission at the base of the FBs is brighter than at high latitudes, with an  energy spectrum that remains hard $\propto E^{-2}$ up to $\sim 1$ TeV~\cite{TheFermi-LAT:2017vmf, Herold:2019pei}.

Deep observations of the FBs near the GC in very-high-energy (VHE, E$\gtrsim$100 GeV) gamma-rays  can provide crucial insights into their origin, {\it eg.}, whether the bubbles were created by an AGN-like burst or by a star-formation activity near the GC. In this contribution we present for the first time new H.E.S.S. observations of the GC region carried out with the Inner Galaxy Survey (IGS) by H.E.S.S. to search for TeV emission at the base of the FBs.

\section{Observations and Data analysis}
\begin{figure*}[!htbp]
\centering
\includegraphics[width=0.49\textwidth]{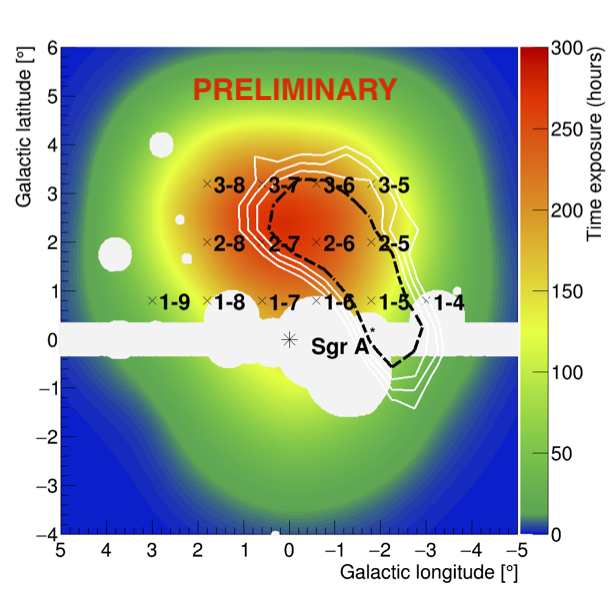}
\caption{Exposure map obtained from the 2014-2020 observations of the Galactic Centre region with H.E.S.S. The H.E.S.S. pointing positions of the Inner Galaxy Survey are marked as black crosses. The white contours show the surface brightness (in sr$^{-1}$) at the base of the FBs obtained from {\it Fermi}-LAT observations. The region of interest for H.E.S.S. is shown as a long-dashed black line. The grey-shaded area corresponds to the set of masks used for the excluded regions.}
\label{fig:skymap}      
\end{figure*}

New observations of the GC region have been performed 
with H.E.S.S. to survey positive Galactic latitudes up to about 6$^\circ$ via the Inner Galaxy Survey (IGS) programme. The dataset analyzed here is obtained from observation runs carried out under dark-sky conditions between 2014 and 2020 with the H.E.S.S. five-telescope array. The telescope pointing positions of the current IGS are shown in Fig.~\ref{fig:skymap}. The observations are taken at zenith angle below 40$^\circ$ with all the telescopes pointings on a set of defined sky positions.
After data quality selection procedure, the dataset amounts to a total of 546 hours of high-quality observations. 

The signal in the H.E.S.S. dataset is searched in a region of interest (ROI), which is hereafter referred to as the ON region, chosen according to the surface brightness of the emission at the base of the FBs. It reaches its maximum  at Galactic longitude $l\approx$ -1$^\circ$ and latitude $b\approx$ 2$^\circ$. The H.E.S.S. ROI defined as the region of the sky in which the FB surface brightness 
is larger than 8.5 sr$^{-1}$. The H.E.S.S. ROI is shown in Fig.~\ref{fig:skymap} together with the FB surface brightness contours. The background is measured according to the reflected background method on a run-by-run basis: the background events are measured in a symmetric region of the ROI with respect to the pointing position of the observational run,  which is hereafter referred to as the OFF region. A set of masks is used as excluded regions in the analysis in order to avoid any contamination of the nearby VHE sources in the signal and background regions.
The signal and background regions have the same solid angle and acceptance and no further normalisation is needed. 

The energy count distributions in the ON and the OFF regions are computed 
together with the excess significance in the spatial bins of the H.E.S.S. ROI 
\begin{figure}[!htbp]
\centering
\includegraphics[width=0.49\textwidth]{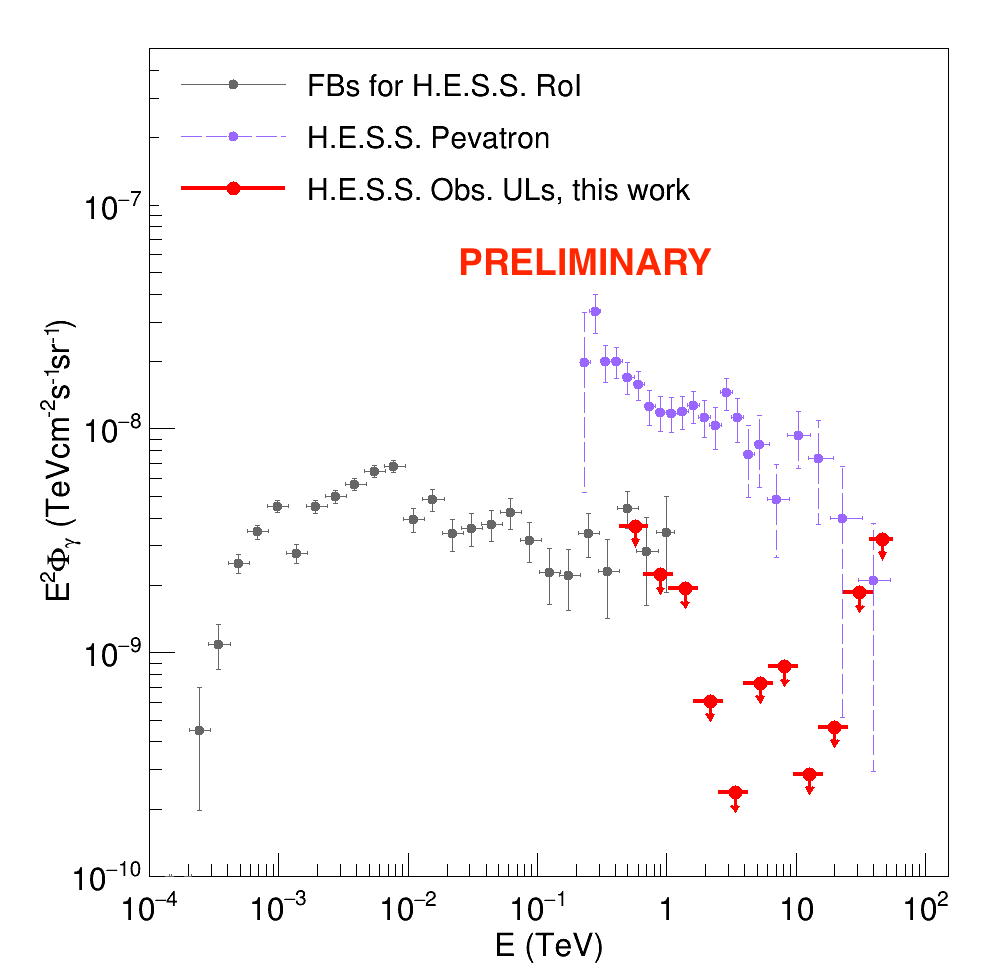}
\caption{Spectral energy distribution. The energy flux measured by {\it Fermi}-LAT in the H.E.S.S. ROI is shown as gray points. The error bars show the 1$\sigma$ statistical uncertainty. 
Observed upper limits computed at 95\% C.L. from H.E.S.S. observations are shown as red arrows.
The solid-angle-averaged flux from the 
H.E.S.S. Pevatron~\cite{Abramowski:2016mir} is also plotted (purple points).}
\label{fig:upperlimit}      
\end{figure}
following the statistical approach of Ref.~\cite{1983ApJ...272..317L}. No significant overall gamma-ray excess is found in the spatial bins of the ROI. Energy differential flux upper limits in 0.2 dex energy bins are computed at 95\% C.L. assuming the best-fit power law index ($\Gamma=1.9$) derived from the {\it Fermi}-LAT data  analysis~\cite{TheFermi-LAT:2017vmf}. 
The  differential flux upper limit is computed according to Ref.~\cite{Rolke:2004mj} assuming a 20\% systematic uncertainty.
Fig.~\ref{fig:upperlimit} shows the energy-differential flux upper limits derived by H.E.S.S. together with the FBs differential flux in the H.E.S.S. ROI. The 95\% C.L. flux upper limit at 1 TeV reaches about $\rm 2\times10^{-9} \,TeVcm^{-2}s^{-1}sr^{-1}$.

\section{Discussion and conclusion}
The high-energy gamma-ray emission measured by {\it Fermi}-LAT can be explained by relativistic electrons scattering off the ambient interstellar radiation field via inverse Compton process, or by inelastic collisions of relativistic protons in the interstellar medium via $pp$ interaction.

For the joint analysis of {\it Fermi}-LAT and H.E.S.S. datasets,  we consider {\it Fermi}-LAT results presented in Ref.~\cite{TheFermi-LAT:2017vmf} at energies $>10$~GeV. The {\it Fermi}-LAT data at energies between 1 and 10 GeV was used to produce the spatial template for the FBs, which can lead to a bias in the FB spectrum. During the analysis we performed joint fit of {\it Fermi}-LAT and H.E.S.S. data assuming a (super)exponential cutoff power-law (SEPL) as well as broken power-law (BPL) gamma-ray spectra. Given the present H.E.S.S. photon statistics, a SEPL spectrum cannot be significantly preferred over a EPL spectrum. The upper limits derived from H.E.S.S. observations on the base of FB 
\begin{figure}[!hb]
\centering
\includegraphics[width=0.55\textwidth]{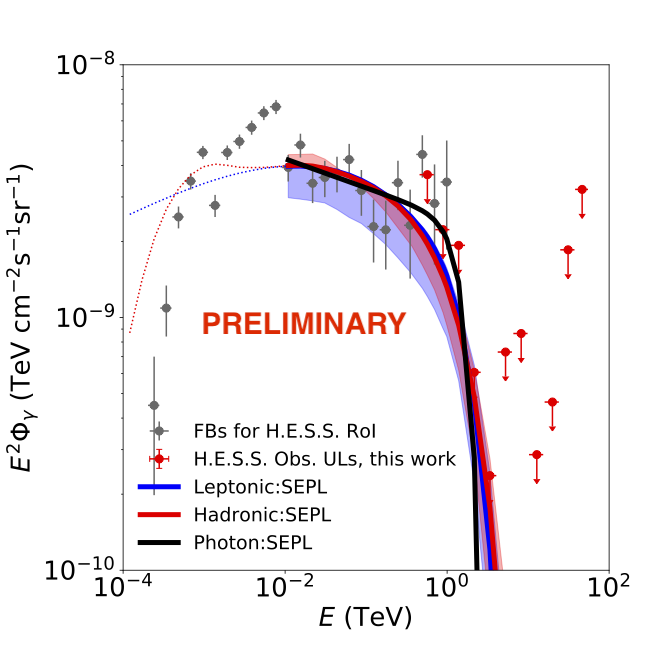}
\caption{Best-fit gamma-ray models to {\it Fermi}-LAT ($>10$~GeV, gray points) and H.E.S.S. datasets assuming a SEPL spectrum for the injected electrons (blue line) and protons (red line), respectively. The error bars show the 1$\sigma$ statistical uncertainty. 
The blue-shaded and red-shaded bands correspond to the 
1$\sigma$ statistical error of the best-fit spectrum.
The energy-differential flux from {\it Fermi}-LAT (black dots) is shown together with H.E.S.S. upper limits
(red arrows).}
\label{fig:constraints}      
\end{figure}
emission constrain the observed cutoff in the photon spectrum to be E$_{\rm \gamma, cut}= 1.1^{+0.6}_{-0.4}$ TeV for exponential cut-off power-law (EPL) spectrum. The 95\% C.L. upper limit on photon energy cutoff assuming the EPL spectrum parametrization is 2.2 TeV. 
Assuming gamma-ray spectra produced within simple one-zone leptonic and hadronic models, we can further constrain the spectral parameters of the parent particle population. The 95\% C.L. upper limit on the energy cutoff of the electron and proton particle spectra are $E_{\rm e,cut}$ = 9.7 TeV and $E_{\rm p,cut}$ = 22.9 TeV, respectively.

The new H.E.S.S. observations of the GC region carried out under the IGS programme provide a new probe of the underlying models responsible of the measured emission by {\it Fermi}-LAT. The present H.E.S.S. dataset constrains the spectral characteristics of leptonic and hadronic models intended to explain the FBs, and could provide further insights on their origin. 

\section{Acknowledgements}
H.E.S.S. gratefully acknowledges financial support from the agencies and organizations listed at: \href{https://www.mpi-hd.mpg.de/hfm/HESS/pages/publications/auxiliary/HESS-Acknowledgements-2021.html}{https://www.mpi-hd.mpg.de/hfm/HESS/pages/publications/auxiliary/HESS-Acknowledgements-2021.html}.

%
%
%

\section*{Full author list: H.E.S.S. Collaboration}
H.~Abdalla$^{1}$, 
F.~Aharonian$^{2,3,4}$, 
F.~Ait~Benkhali$^{3}$, 
E.O.~Ang\"uner$^{5}$, 
C.~Arcaro$^{6}$, 
C.~Armand$^{7}$, 
T.~Armstrong$^{8}$, 
H.~Ashkar$^{9}$, 
M.~Backes$^{1,6}$, 
V.~Baghmanyan$^{10}$
% ...

\medskip

\noindent
$^{1}$University of Namibia, Department of Physics, Private Bag 13301, Windhoek 10005, Namibia\\
$^{2}$Dublin Institute for Advanced Studies, 31 Fitzwilliam Place, Dublin 2, Ireland\\
$^{3}$Max-Planck-Institut f\"ur Kernphysik, P.O. Box 103980, D 69029 Heidelberg, Germany\\
$^{4}$High Energy Astrophysics Laboratory, RAU,  123 Hovsep Emin St  Yerevan 0051, Armenia\\
$^{5}$Aix Marseille Universit\'e, CNRS/IN2P3, CPPM, Marseille, France\\
$^{6}$Centre for Space Research, North-West University, Potchefstroom 2520, South Africa\\
$^{7}$Laboratoire d'Annecy de Physique des Particules, Univ. Grenoble Alpes, Univ. Savoie Mont Blanc, CNRS, LAPP, 74000 Annecy, France\\
$^{8}$University of Oxford, Department of Physics, Denys Wilkinson Building, Keble Road, Oxford OX1 3RH, UK\\
$^{9}$IRFU, CEA, Universit\'e Paris-Saclay, F-91191 Gif-sur-Yvette, France\\
$^{10}$Instytut Fizyki J\c{a}drowej PAN, ul. Radzikowskiego 152, 31-342 Krak{\'o}w, Poland\\
% ...
\end{document}